\documentclass[twocolumn,showpacs,preprintnumbers,amsmath,amssymb]{revtex4}

\usepackage{graphicx}
\usepackage{dcolumn}
\usepackage{bm}

\newcommand{\fracpartial}[2]{\frac{\partial #1}{\partial #2}}

\newcommand{\innerP}[2]{\mbox{$\left(#1 \cdot #2\right)_{\perp}$}}
\newcommand{\braketo}[3]{\mbox{$\langle #1 | #2 | #3 \rangle$}}

\newcommand{\ket}[1]{\mbox{$| #1 \rangle$}}

\begin{document}

\preprint{}

\title{Nonrigid chiral soliton for the octet and decuplet baryons}

\author{Satoru Akiyama}
\email{akiyama@ph.noda.tus.ac.jp}
\author{Yasuhiko Futami}
\affiliation{
	Department of Physics, Tokyo University of Science,
	2641, Noda, Chiba 278-8510, Japan 
}

\date{\today}

\begin{abstract}
Systematic treatment of the collective rotation of the nonrigid chiral soliton 
is developed in the SU(3) chiral quark soliton model
and applied to the octet and decuplet baryons.
The strangeness degrees of freedom are treated by a simplified bound-state approach
which omits the locality of the kaon wave function.
Then, the flavor rotation is divided into the isospin rotation and
the emission and absorption of the kaon.
The kaon Hamiltonian is diagonalized by the Hartree approximation.
The soliton changes the shape according to the strangeness.
The baryons appear as the rotational bands of the combined system of the soliton and the kaon.
\end{abstract}


\maketitle

\section{Introduction\label{sec:intro}}
The soliton picture of the baryon is employed by many effective models of QCD
in the low energy region: the Skyrme model
\cite{rf:Skyrme61,rf:Adkin83,rf:Guadanini84,rf:Mazur84,rf:Yabu88,rf:Balachandran85},
the Nambu--Jona-Lasinio model \cite{rf:Nambu61,rf:Ebert86,rf:Weigel92}
and the chiral quark soliton model (CQSM)
\cite{rf:Diakonov88,rf:Reinhardt88,rf:Reinhardt89,rf:Wakamatsu91}.
A technical feature of the picture is the separation of the internal and
external degrees of freedom.
The internal degrees of freedom (mesons, quarks) nonlinearly interact with each other and
construct the soliton self-consistently.
On the other hand, the external degrees of freedom describe the collective motions
of the soliton: rotation, vibration, translational motion. 
To solve the coupled equations for these degrees of freedom, for technical reason,
one should assume the decoupling of the internal degrees of freedom with the collective motion
and employ the simplest ansatz for the soliton: the hedgehog shape.
For example, in the CQSM, the equations of motion for the soliton consist of the local
functions of the meson profile and the infinite eigenstates of the quarks
in the background soliton. Then, the equations should be self-consistently solved.
The process requires a large number of calculations by the computer.
And it is hard to self-consistently incorporate the interaction between the nonrigid soliton
and the collective motion.
Thus, the symmetry of the soliton is a technically important ingredient
to reduce the number of calculational steps. 
From the physical point of view, however, it seems oversimplified that
one assumes the symmetry which is broken by the collective rotation.

In the previous work \cite{rf:akiyama03}, we studied these assumption
for the octet baryons with the CQSM.
To overcome the complexities of the self-consistent calculation for the nonrigid soliton
and incorporate the interaction between the soliton and the collective motion,
we introduced the physical and/or mathematical tools:
the mean field approximation for the rotated system,
the tensor operator $Z^{JLSI}_{KK_{3}}({\bf \hat{r}})$ for the grand spin
which is used to expand the chiral field, and
the explicit form of the isospin vector of the chiral soliton with the quark states.
The result indicated that the hedgehog shape of the soliton is stable only for the nucleon ($N$)
and unstable for the strange baryons ($\Lambda$,$\Sigma$,$\Xi$).
Thus, the soliton changes the shape according to the strangeness of the baryon.

Although these tools are applicable to the more general systems,
the collective Hamiltonian takes the complicated form and
the outlook of the collective quantization is bad due to the less symmetry 
of the nonrigid soliton than the hedgehog one.
Therefore, in this article, we reformulate the collective Hamiltonian and improve
the procedure of the collective quantization given in Ref.~\cite{rf:akiyama03}.
Furthermore, we introduce the Hartree approximation for the kaon to simplify
the treatment of the particle-antiparticle creation \cite{rf:akiyama03}.
We apply these improved tools to the octet and decuplet baryons.

In Sec.~\ref{sec:model}, we review the SU(3) CQSM and the mean field approximation
for the rotated system.
In Sec.~\ref{sec:var_tr}, the variable transformation between the two types
of the parameterization for the flavor rotation is given.
In Sec.~\ref{sec:coll_q}, first, we construct the collective Hamiltonian using the variable
transformation.
Next, we show the Hartree approximation for the kaon in the background soliton and
define the basis for the collective quantization using the Hartree states.
In Sec.~\ref{sec:results}, we show the numerical results of the Hartree approximation
and the collective quantization.
Finally, in Sec.~\ref{sec:summary} we summarize the results and discuss the relation
between the deformation of the chiral soliton and the kaon.

\section{Model and the mean field approximation for the flavor rotation \label{sec:model}}
Hereafter we follow the notation used in Ref.~\cite{rf:akiyama03} for the various quantities.
The chiral quark soliton model in the case of flavor SU(3) is given
by the path integral with respect to the chiral meson fields and
the quark fields \cite{rf:Diakonov88,rf:Reinhardt88,rf:Blotz93}.
We postulate the so called cranking form \cite{rf:Adkin83,rf:Braaten88}
for the meson field:
\begin{equation}
	U^{\gamma_{5}}(x) = {\cal A}(t)B^{\dagger}(t)
		U^{\gamma_{5}}_{0}({\bf r}) B(t){\cal A}^{\dagger}(t),
\label{eq:aua}
\end{equation}
where $U^{\gamma_{5}}_{0}({\bf r})$ is the static meson field,
${\cal A}(t)$ describes the adiabatic rotation of the system in SU(3) flavor space,
and $B(t)$ describes the spatial rotation.
Then, the effective action for $U^{\gamma_{5}}$ is reduced to 
\begin{equation}
	i S_{F} = N_{c} 
		\log\det \left(
			i\partial_{t}
			+i {\cal A}^{\dagger}\dot{\cal A}
			+i B\dot{B}^{\dagger}
			-H'
		\right),
\label{eq:action}
\end{equation}
where $\det$ denotes the functional determinant for the quarks,
$N_{c}$ is the number of colors, and $H'$ is the rotated quark Hamiltonian
defined below.

For $U^{\gamma_{5}}_{0}$, we assume the embedding of the SU(2) field
to the SU(3) matrix but do not assume the hedgehog shape \cite{rf:akiyama03}.

We write the flavor rotation \cite{rf:Kaplan90} as
\begin{equation}
	{\cal A}(t) = \left(\begin{array}{cc}
					A(t)		& 0\\
					0^{\dagger} & 1
			\end{array}\right) A_{s}(t),
\label{eq:A_param}
\end{equation}
where $A$ is the flavor SU(2) rotation operator and 
$A_{s}$ represents the rotation into the strange directions.
In particular, we parameterize $A_{s}(t)$ as
\begin{equation}
	A_{s}(t) = \exp i \left(
				\begin{array}{cc}
				0 & \sqrt{2} D(t)\\
				\sqrt{2} D^{\dagger}(t) & 0
				\end{array}
			\right),
\label{eq:As_param}
\end{equation}
where $D = (D_{1}, D_{2})^{T}$ is the isodoublet spinor.
From the transformation property under the flavor rotation, we call $D$ the kaon field.
In Ref.~\cite{rf:Westerberg94}, it is argued that $D \sim 1/\sqrt{N_{c}}$
in the large $N_{c}$ limit due to the Wess-Zumino term,
even if the strange quark mass is light.
We also employ the classification and treat $D$ perturbatively. 
Although Eq.~(\ref{eq:As_param}) was motivated by the bound-state approach \cite{rf:Callan85},
we will not treat the locality of the kaon wave function \cite{rf:Westerberg94} in this article.
Inclusion of the locality is a complicated task.

From the current quark mass matrix:
$\hat{m} = m_{0} \lambda_{0} + m_{8} \lambda_{8}$ ($m_{3}=0$),
we obtain the rotated one:
\begin{equation}
	\hat{m}' = {\cal A}^{\dagger} \hat{m} {\cal A}
	= m_{0} \lambda_{0} + m_{8} D^{(8)}_{8\mu}(A_{s}) \lambda_{\mu},
\label{eq:rot_mq}
\end{equation}
where $D^{(8)}_{\mu\nu}(A_{s})$ ($\mu,\nu = 1,2,\dots,8$) is the Wigner \textit{D}
function of $A_{s}$ in the adjoint representation:
\begin{equation}
	D^{(8)}_{\mu\nu}(A_{s}) = \frac{1}{2} {\rm tr} \left(
			A_{s}^{\dagger} \lambda_{\mu} A_{s} \lambda_{\nu}
		\right).
\label{eq:d_func}
\end{equation}
The value of $m_{8}$ represents the strength of the flavor SU(3) symmetry breaking. 

Here, we define the following quantities:  
\begin{eqnarray}
	\kappa_{0} &\equiv& 2D^{\dagger} D,\\
	\kappa_{j} &\equiv& 2D^{\dagger} \tau_{j} D,
\end{eqnarray}
where $j = 1,2,3$, and give the explicit form of $D^{(8)}_{\mu\nu}(A_{s})$
in the Appendix~\ref{sec:rot_s}.
Suppose that the collective motions $\cal A$ and $B$ are quantized
and $\ket{B}$ as an eigenstate of the collective Hamiltonian.
If $\ket{B}$ points to a specific direction in the isospin space,
\begin{eqnarray}
 	\kappa_{B0} &=& \braketo{B}{\kappa_{0}}{B},
\label{eq:kappa0}\\
	\kappa_{B3} &=& \braketo{B}{\kappa_{3}}{B}
\label{eq:kappa3}
\end{eqnarray}
have nonzero values \cite{rf:akiyama03}. 
Then, the expectation value $\braketo{B}{\hat{m}'}{B}$ may be approximated by
\begin{equation}
	\hat{m}_{B} = m_{0} \lambda_{0}
		+ m_{B3} \lambda_{3} + m_{B8} \lambda_{8},
\label{eq:mq_eff}
\end{equation}
with
\begin{equation}
	m_{B\mu} = 
	m_{8} \lim_{\kappa_{0,3} \rightarrow \kappa_{B0,3}}{D^{(8)}_{8\mu}(A_{s})}
	\hspace{5mm}(\mu = 3,8).\\ 
\label{eq:mqT_eff}
\end{equation}
We call $\hat{m}_{B}$ the mean field value of $\hat{m}'$.

Using these quantities, the rotated quark Hamiltonian $H'$
in Eq.~(\ref{eq:action}) is given by
\begin{eqnarray}
	H' &=& H'_{0} + \Delta H',
\label{eq:rot_hq}\\
	H'_{0} &=& \frac{1}{i} {\bm \alpha} \cdot \nabla + 
		\beta \left(MU^{\gamma_{5}}_{0}+\hat{m}_{B}\right),
\label{eq:rot_hq0}\\
	\Delta H' &=& \beta \left(\hat{m}' - \hat{m}_{B}\right)
		= \beta T_{\mu} \sigma_{\mu},
\label{eq:rot_hq1}
\end{eqnarray}
where $M$ is the dynamical quark mass,
$\sigma_{\mu}$ ($\mu = 1,2,\dots,8$) defines the fluctuation
around the mean field, and $T_{\mu} = \lambda_{\mu}/2$.
$H'_{0}$ contain the effects of the flavor SU(3) and SU(2) symmetry breaking
through the mean field $\hat{m}_{B}$. 

We expand $S_{F}$ in power of the angular velocities
($i {\cal A}^{\dagger}\dot{\cal A}$, $i B\dot{B}^{\dagger}$)
and $\Delta H'$ around the eigenstate of $(i\partial_{t}-H'_{0})$ 
to get the effective Lagrangian $\cal L$ \cite{rf:akiyama03,rf:Reinhardt89}.
The former corresponds to the $1/N_{c}$ expansion
\cite{rf:tHooft74,rf:Witten79,rf:Dashen94}.
We retain the order of the expansion up to $O(1/N_{c})$. 
The latter corresponds to the perturbation in power of $\sigma_{\mu}$,
which is given by the product of $m_{8}$ and the fluctuation
of $D_{8\mu}^{(8)}(A_{s})$ around the expectation value.
We assume that the large parts of the flavor SU(3) symmetry breaking
are included in the lowest order term of $S_{F}$ through the mean field $\hat{m}_{B}$ 
and the residual effects can be estimated by the first order perturbation in $\sigma_{\mu}$.
In fact, the fluctuations is small, since the expectation values of $D_{8\mu}^{(8)}$
are evaluated according to the individual baryon states.
For example, the fluctuation of $D^{(8)}_{88}$ starts from the difference between
the $O(1/N_{c})$ quantities in the large $N_{c}$ limit: $-3(\kappa_{0}-\kappa_{B0})/2$.
Furthermore, the fluctuation is always accompanied with $m_{8}$.
The value of $m_{8}$ is about $200$~MeV and fairly small
compared with the typical energy scale $\Lambda$
which is the energy cutoff parameter around $700$~MeV.
Here, the small value of $m_{8}/\Lambda$ does not contradict
the perturbative treatment of $D$ in the large $N_{c}$ limit,
since the kaon field $D \sim 1/\sqrt{N_{c}}$ independent of the strange quark mass,
as noted above.

The static meson field $U^{\gamma_{5}}_{0}$ is self-consistently determined
\cite{rf:Reinhardt88} by a variation of the classical soliton energy $E_{cl}$
which corresponds to the lowest order term of $S_{F}$.
Because of Eq.~(\ref{eq:mq_eff}), it introduces the isospin symmetry breaking
$\kappa_{B3} \not= 0$ to the system in the body fixed frame.
Here, $\kappa_{B3} \not= 0$ leads to an axially symmetric deformation
of the chiral soliton \cite{rf:akiyama03}.

It was pointed out in Ref.~\cite{rf:Dorey94} that the Skyrmion must be
improved by the flavor SU(2) rotation to resolve the "Yukawa problem."
Then, the shape of the Skyrmion deviates from the hedgehog one in the body fixed frame.
The hedgehog shape of the soliton is the leading order term in the large $N_{c}$ limit
and the deviations correspond to the higher order terms.
In our case, the effects of the rotation into the strangeness directions
deform the hedgehog soliton.

\section{Variable Transformation \label{sec:var_tr}}
We argue the relation between the two types of the parameterization
for the flavor rotation. 
First, we define the local variables $\alpha$ and $a$
of the rotation ${\cal A}$ and $A$, respectively by
\begin{eqnarray}
	\dot{\alpha}^{\mu} T_{\mu} &=& -i {\cal A}^{\dagger}\dot{\cal A}\nonumber\\
		&=& A_{s}^{\dagger} \left(\begin{array}{cc}
					- i A^{\dagger}\dot{A} & 0\\
					0^{\dagger} & 0
			\end{array}\right) A_{s} - i A_{s}^{\dagger}\dot{A_{s}},
\label{eq:local_su3}\\
	\dot{a}^{j} \frac{\tau_{j}}{2} &=& -i A^{\dagger} \dot{A},
\label{eq:local_su2}
\end{eqnarray}
where $\mu = 1,2,\dots,8$, and $j = 1,2,3$.
Among $\dot{\alpha}^{\mu}$, $\dot{a}^{j}$, and $\dot{D}$, there is a relation: 
\begin{equation}
	\dot{\alpha}^{\mu} = \dot{a}^{j} D^{(8)}_{j\mu}(A_{s})
				+ \dot{D}^{\dagger} C_{\mu}(A_{s})
				+ C^{\dagger}_{\mu}(A_{s}) \dot{D},
\label{eq:v_rel}
\end{equation}
where $C_{\mu}$ is a isodoublet spinor defined by
\begin{equation}
	C_{\mu}(A_{s}) \equiv \frac{1}{i} {\rm tr} \left(
			A_{s}^{\dagger}\fracpartial{A_{s}}{D^{\dagger}} \lambda_{\mu}
		\right),
\label{eq:c_cof}
\end{equation}
and its explicit form is given in the Appendix~\ref{sec:rot_s}. 
Spinors $C_{\mu}$ and $D$ are related to the Wigner \textit{D} function
$D^{(8)}_{\mu\nu}(A_{s})$ by the following equations:
\begin{eqnarray}
	i \left(
		D^{\dagger} \frac{\tau_{j}}{2} C_{\mu}
		-C^{\dagger}_{\mu} \frac{\tau_{j}}{2} D
	\right)
	&=& \delta_{j\mu} - D^{(8)}_{j\mu},
\label{eq:dc_product_t}\\
	i \left(
		D^{\dagger} C_{\mu} - C^{\dagger}_{\mu} D
	\right)
	&=& \frac{2}{\sqrt{3}} \left(\delta_{8\mu} - D^{(8)}_{8\mu} \right).
\label{eq:dc_product_s}
\end{eqnarray}

Next, we investigate the relation among the canonical momenta.
The effective Lagrangian $\cal L$ which is a function of $\dot{\alpha}^{\mu}$,
is also a function of $\dot{a}^{j}$ and $\dot{D}$ through Eq.~(\ref{eq:v_rel}). 
The canonical momenta conjugate to $\alpha^{\mu}$ are defined by
\begin{equation}
	I^{\cal A}_{\mu} = \fracpartial{\cal{L}}{\dot{\alpha}^{\mu}}.
\label{eq:can_IA}
\end{equation}
On the other hand, the canonical momenta conjugate to $a^{j}$ and $D$ are defined by
\begin{eqnarray}
	I_{j} &=& \fracpartial{\cal{L}}{\dot{a}^{j}} = D^{(8)}_{j\mu} I^{\cal A}_{\mu},
\label{eq:can_I}\\
	P &=& \fracpartial{\cal{L}}{\dot{D}^{\dagger}} = C_{\mu} I^{\cal A}_{\mu}.
\label{eq:can_p}
\end{eqnarray}
For the isospin rotation,
using Eqs.~(\ref{eq:dc_product_t}),(\ref{eq:can_I}) and (\ref{eq:can_p}), we obtain
\begin{equation}
	I^{\cal A}_{j} = I_{j} + I_{Kj},
\label{eq:isospin_rel}
\end{equation}
where $I_{Kj}$ is the isospin carried by the kaon:
\begin{equation}
	I_{Kj} = i\left(
		D^{\dagger} \frac{\tau_{j}}{2} P - P^{\dagger} \frac{\tau_{j}}{2} D
	\right).
\label{eq:kaon_ispin}
\end{equation}
Furthermore, using Eq.~(\ref{eq:dc_product_t}), we obtain
\begin{equation}
	Y = S - \frac{2}{\sqrt{3}} I^{\cal A}_{8},
\end{equation}
where the term $-(2/\sqrt{3}) I^{\cal A}_{8}$ is the baryon number \cite{rf:Rabinovic84},
$S$ is the strangeness carried by kaon, and $Y$ is the (left) hypercharge defined by
\begin{eqnarray}
	S &=& i \left(D^{\dagger} P-P^{\dagger} D\right),
\label{eq:kaon_s}\\
	Y &=& - \frac{2}{\sqrt{3}} D^{(8)}_{8\mu} I^{\cal A}_{\mu}.
\label{eq:hyper_charge}
\end{eqnarray}

We define
\begin{equation}
	\tilde{P} = P - C_{j} I^{\cal A}_{j} - C_{8} I^{\cal A}_{8}.
\end{equation}
For the rotation into the strange direction,
Eq.~(\ref{eq:can_p}) is rewritten as 
\begin{equation}
	C_{a} I^{\cal A}_{a} = \tilde{P}.
\end{equation}
where $a = 4,5,6,7$.
This equation is solved for $I^{\cal A}_{a}$ as follows.
\begin{eqnarray}
	I^{\cal A}_{4} &=& -\frac{\sqrt{3}}{2 N_{1}}
			\left(D^{(8)}_{84} F_{1} + D^{(8)}_{85} G_{1}\right),\nonumber\\
	I^{\cal A}_{5} &=& -\frac{\sqrt{3}}{2 N_{1}}
			\left(D^{(8)}_{85} F_{1} - D^{(8)}_{84} G_{1}\right),\nonumber\\
	I^{\cal A}_{6} &=& -\frac{\sqrt{3}}{2 N_{2}}
			\left(D^{(8)}_{86} F_{2} + D^{(8)}_{87} G_{2}\right),\nonumber\\
	I^{\cal A}_{7} &=& -\frac{\sqrt{3}}{2 N_{2}}
			\left(D^{(8)}_{87} F_{2} - D^{(8)}_{86} G_{2}\right),
\label{eq:kaon_rel}
\end{eqnarray}
where $N_{i}$,$F_{i}$, and $G_{i}$ ($i = 1,2$) are functions of $D$
and $\tilde{P}$, and given in the Appendix~\ref{sec:rot_s}.

\section{Collective Quantization \label{sec:coll_q}}
In the same manner as Eq.~(\ref{eq:can_IA}), we introduce the canonical momenta $J_{j}$
conjugate to the local variables $b^{j}$ of $B(t)$ \cite{rf:akiyama03,rf:Braaten88}.
Then, the collective Hamiltonian is given by
\begin{widetext}
\begin{eqnarray}
	{\cal H} &=& \dot{\alpha}^{\mu} I^{\cal A}_{\mu} + \dot{b}^{j} {J}_{j}
			- {\cal L}\nonumber\\
	&=& E_{cl}
	+\frac{1}{2 U_{33}} \left(J_{3}-B_{3}
		+\sigma_{3} \Delta_{33}+\sigma_{8} \Delta_{38}\right)^{2}
	+\sigma_{3} \Gamma_{3}+\sigma_{8} \Gamma_{8}\nonumber\\
	&&+\frac{1}{2 \left(U_{11}V_{11}-W_{11}^{2}\right)}
	\sum_{j=1}^{2} [
		V_{11} \left(I^{\cal A}_{j}-\sigma_{j} \Delta_{11}\right)^{2}
		+U_{11} \left(J_{j}+\sigma_{j} \tilde{\Delta}_{11}\right)^{2} \nonumber\\
		&&+2 W_{11} \left(I^{\cal A}_{j}-\sigma_{j} \Delta_{11}\right)
			\left(J_{j}+\sigma_{j} \tilde{\Delta}_{11}\right)
	]\nonumber\\
	&&+\frac{1}{2 U_{44}}\left[
		\left(I^{\cal A}_{4}-\sigma_{4} \Delta_{44}\right)^{2}
		+\left(I^{\cal A}_{5}-\sigma_{5} \Delta_{44}\right)^{2}
	\right]
	+\frac{1}{2 U_{66}}\left[
		\left(I^{\cal A}_{6}-\sigma_{6} \Delta_{66}\right)^{2}
		+\left(I^{\cal A}_{7}-\sigma_{7} \Delta_{66}\right)^{2}
	\right],
\label{eq:h}
\end{eqnarray}
\end{widetext}
where $E_{cl}$,$U_{\mu\nu}$,$V_{\mu\nu}$,$W_{\mu\nu}$,$\Delta_{\mu\nu}$,
$\tilde{\Delta}_{\mu\nu}$,$B_{\mu}$, and $\Gamma_{\mu}$
are defined in Ref.~\cite{rf:akiyama03} and
calculated with the quark states in the background soliton.
Especially, $E_{cl}$ is the classical soliton energy.
Because of the axial symmetry of the chiral field, there is the following constraint
on the canonical momenta:
\begin{equation}
	I^{\cal A}_{3} + J_{3} = 0.
\label{eq:constraint}
\end{equation}

In our approach, the Hamiltonian and the constraint should be considered
as functions of the variables $a^{j}$,$b^{j}$,$D$,
and the momenta $I_{j}$,$J_{j}$,$P$ with Eqs.~(\ref{eq:isospin_rel}) and (\ref{eq:kaon_rel}).
Furthermore, we expand ${\cal H}$ up to the first order in power of $1/N_{c}$ and $\sigma_{\mu}$.
The explicit form of the collective Hamiltonian is given in Appendix~\ref{sec:h_coll}.

In Ref.~\cite{rf:akiyama03}, the derivation of the Hamiltonian from the Lagrangian
was performed by the perturbation with respect to $1/N_{c}$ and $m_{8}$.
As a result, the Hamiltonian had a highly complicated form.
On the other hand, the present form of $\cal H$ is simple and has a good foresight.  

For the quantization of the system, we separate ${\cal H}$ as
\begin{equation}
	{\cal H} = E_{cl} + {\cal H}_{K} + {\cal H}_{rot} + {\cal H}_{int},
\end{equation}
where ${\cal H}_{K}$ is the part containing only $D$ and $P$ and
describes the kaon in the background soliton,
${\cal H}_{rot}$ is the part containing only $I_{j}$ and $J_{j}$ and
represents the collective rotation of the system
in isospin and real space, and ${\cal H}_{int}$ represents
an interaction between the kaon and the rotation.

\subsection{Kaon Hamiltonian ${\cal H}_{K}$ \label{sec:h_k}}
We separate ${\cal H}_{K}$ into a bilinear part ${\cal H}_{K0}$  
and a higher order part ${\cal H}_{K1}$ in power of the kaon operators $D$ and $P$.
\begin{equation}
	{\cal H}_{K} = {\cal H}_{K0} + {\cal H}_{K1}.
\end{equation}
The explicit form of the these quantities are given in Appendix~\ref{sec:h_coll}.

${\cal H}_{K0}$, $S$ [Eq.~(\ref{eq:kaon_s})], and $I_{K3}$ [Eq.~(\ref{eq:kaon_ispin})] 
can be exactly diagonalized as follows \cite{rf:akiyama03,rf:Westerberg94}.
\begin{eqnarray}
	{\cal H}_{K0} &=& E_{ind}
	+ \sum_{\gamma = 1}^{2} \left(
		\omega_{\gamma} \xi_{\gamma}^{\dagger} \xi_{\gamma}
		+ {\bar \omega}_{\gamma}  \bar{\xi}_{\gamma}^{\dagger} \bar{\xi}_{\gamma}
	\right),
\label{eq:hK0_diag}\\
	S &=& \sum_{\gamma = 1}^{2} \left(
			\xi_{\gamma}^{\dagger} \xi_{\gamma}
			- \bar{\xi}_{\gamma}^{\dagger} \bar{\xi}_{\gamma}
		\right),
\label{eq:s_diag}\\
	I_{K3} &=& \frac{1}{2} \left(
		\xi_{1}^{\dagger} \xi_{1} - \bar{\xi}_{1}^{\dagger} \bar{\xi}_{1}
		- \xi_{2}^{\dagger} \xi_{2} + \bar{\xi}_{2}^{\dagger} \bar{\xi}_{2}
	\right),
\label{eq:ik3_diag}
\end{eqnarray}
where $E_{ind}$ is defined in Appendix~\ref{sec:h_coll},
$\gamma$ is isospin index, $\xi_{\gamma}^{\dagger}$ ($\xi_{\gamma}$) and
$\bar{\xi}_{\gamma}^{\dagger}$ ($\bar{\xi}_{\gamma}$) are
the creation (annihilation) operators for the kaon and antikaon, respectively,
and $\omega_{\gamma}$ (${\bar \omega}_{\gamma}$) is the energy eigenvalue of the kaon (antikaon).
The Fock space is generated by successive operation of the creation operators
on the vacuum state $\ket{0}$:
\begin{equation}
	\ket{n_{1}, \bar{n}_{1}, n_{2}, \bar{n}_{2}} = \prod_{\gamma = 1}^{2}
		\frac{1}{\sqrt{n_{\gamma}!\, \bar{n}_{\gamma}!}}
		\left(\xi_{\gamma}^{\dagger}\right)^{n_{\gamma}}
		\left(\bar{\xi}_{\gamma}^{\dagger}\right)^{\bar{n}_{\gamma}} \ket{0},
\label{eq:fock}
\end{equation}
where $n_{\gamma}$ and $\bar{n}_{\gamma}$ are some positive integers.

Henceforth, any quantity depending on $D$ and $P$ is considered
as a normal ordered operator with respect to
$\xi_{\gamma}^{\dagger}$,$\xi_{\gamma}$,$\bar{\xi}_{\gamma}^{\dagger}$,
and $\bar{\xi}_{\gamma}$.
Using this prescription, we can treat the anharmonic terms of ${\cal H}$
without any ambiguity.
The choice of ${\cal H}_{K0}$ is important, because the creation and annihilation operators
are given by the diagonalization of ${\cal H}_{K0}$.
In Ref.~\cite{rf:akiyama03}, we use the $O(1)$ Hamiltonian ${\cal H}_{0}$
of ${\cal H}$ in the large $N_{c}$ limit to define the creation and annihilation operators.
The difference between ${\cal H}_{K0}$ and ${\cal H}_{0}$
is higher order terms in the large $N_{c}$ limit.
In the previous work, we classified $B_{3}$ as an $O(1/N_{c})$ quantity and
treated these terms in the rotation-kaon interaction.
In this article, we throw this classification away and require only
that ${\cal H}_{K0}$ contains all bilinear term with respect to $D$ and $P$ in $\cal H$.

To incorporate the effects of the anharmonic term ${\cal H}_{K1}$,
we employ the Hartree approximation.
In this approximation, a bare isodoublet spinor
\begin{equation}
	X_{\gamma}^{\dagger} = \left(\begin{array}{c}
				\xi_{\gamma}^{\dagger}\\
				\bar{\xi}_{\gamma}
	\end{array}\right),
\end{equation}
is transformed to a dressed one $\tilde{X}_{\gamma}^{\dagger}$ by a unitary transformation:
\begin{equation}
	\tilde{X}_{\gamma}^{\dagger} = e^{i G} X_{\gamma}^{\dagger} e^{-i G},
\label{eq:unitary_tr}
\end{equation}
where $G$ is a Hermitian operator.
Since the kaon Hamiltonian ${\cal H}_{K}$ commutes with $S$ and $I_{K3}$, 
$G$ takes the following form:
\begin{equation}
	G = \sum_{\gamma=1}^{2} X^{\dagger}_{\gamma} g_{\gamma} X_{\gamma},
\label{eq:G_hermite}
\end{equation}
where $g_{\gamma}$ are $2 \times 2$ Hermite matrices.

The unitary transformation Eq.~(\ref{eq:unitary_tr}) with Eq.~(\ref{eq:G_hermite})
does not change $S$ and $I_{K3}$.
The dressed state with the good quantum numbers $S$ and $I_{K3}$ is defined as
\begin{equation}
	\ket{B(S, I_{K3})} = e^{i G} \ket{n_{1}, \bar{n}_{1}, n_{2}, \bar{n}_{2}},
\end{equation}
where $S = n_{1}-\bar{n}_{1}+n_{2}-\bar{n}_{2}$ and 
$I_{K3} = (n_{1}-\bar{n}_{1}-n_{2}+\bar{n}_{2})/2$.
We assume that the bare state $\ket{n_{1}, \bar{n}_{1}, n_{2}, \bar{n}_{2}}$ contains
only the valence (anti)kaons \cite{rf:Westerberg94} and the kaon-antikaon pairs are created
by the unitary transformation Eq.~(\ref{eq:unitary_tr}) with
\begin{eqnarray}
	g_{\gamma} &=& \left(\begin{array}{cc}
			0 & f_{\gamma}\\
			f_{\gamma}^{*} & 0
		\end{array}\right),
\label{eq:g_hermite}
\end{eqnarray}
where $f_{\gamma}$ is a complex constant.
Since $S \leq 0$ for the octet and decuplet baryons,
\begin{equation}
	\ket{B(S, I_{K3})} = e^{i G} \ket{0, \bar{n}_{1}, 0, \bar{n}_{2}}.
\label{eq:dressed}
\end{equation}
Here $S = -\bar{n}_{1}-\bar{n}_{2}$ and $I_{K3} = (-\bar{n}_{1}+\bar{n}_{2})/2$.
For the fixed $S$, we obtain the "multiplet" $\ket{B(S, I_{K3})}$ obeying
\begin{equation}
	-S \geq 2I_{K3} \geq S,
\end{equation}
since $\bar{n}_{1},\bar{n}_{2} \geq 0$.
The energy eigenvalue of the soliton $+$ the kaon system is calculated by
\begin{equation}
	E_{B}(S, I_{K3}) = E_{cl} + \braketo{B(S, I_{K3})}{{\cal H}_{K}}{B(S, I_{K3})}.
\label{eq:e_b}
\end{equation}
Since the isospin symmetry is broken by the soliton field \cite{rf:akiyama03},
the states $\ket{B(S, I_{K3})}$ with the different $I_{K3}$ are not degenerate
even within the multiplet.
Hereafter, the lowest eigenvalue is denoted by $E_{B0}(S, I_{K3})$ and 
the corresponding state is denoted by $\ket{B_{0}(S, I_{K3})}$.

The matrices $g_{\gamma}$ are determined by the variational equations
for the lowest state $\ket{B_{0}(S, I_{K3})}$:
\begin{eqnarray}
	\frac{\delta \braketo{B_{0}(S, I_{K3})}{{\cal H}_{K}}{B_{0}(S, I_{K3})}}{\delta g_{\gamma}} = 0.
\label{eq:var_g}
\end{eqnarray}
Thus, the values of $g_{\gamma}$ are given for the individual $\ket{B_{0}(S, I_{K3})}$
with $S$ corresponding to the baryons, and used for calculation of the higher states
$\ket{B(S, I_{K3})}$ in the same multiplet.

Furthermore, we demand that the values of $\kappa_{B0}$ and $\kappa_{B3}$
[Eqs.~(\ref{eq:kappa0}) and (\ref{eq:kappa3})] are determined
by the self-consistent equations for $\ket{B_{0}(S, I_{K3})}$:
\begin{eqnarray}
 	\kappa_{B0} &=& \braketo{B_{0}(S, I_{K3})}{\kappa_{0}}{B_{0}(S, I_{K3})},
\label{eq:kappaB0}\\
	\kappa_{B3} &=& \braketo{B_{0}(S, I_{K3})}{\kappa_{3}}{B_{0}(S, I_{K3})}.
\label{eq:kappaB3}
\end{eqnarray} 
We call the state $\ket{B_{0}(S, I_{K3})}$ with Eqs.~(\ref{eq:var_g}),(\ref{eq:kappaB0}),
and (\ref{eq:kappaB3}) the Hartree state.

Here, one should recognize that the Hartree state does not necessarily correspond 
to the absolute minimum of $E_{B0}(S, I_{K3})$ in the parameter space
$(\kappa_{B0},\kappa_{B3})$.
It is because that the solution $\kappa_{B0}$ of Eq.~(\ref{eq:kappaB0})
is restricted by the strangeness of the baryon and
the absolute minimum can be unphysical point which does not correspond to the solution.

\subsection{Collective rotation Hamiltonian ${\cal H}_{rot}$
and kaon-rotation interaction Hamiltonian ${\cal H}_{int}$ \label{sec:h_rot}}
The explicit forms of ${\cal H}_{rot}$ and ${\cal H}_{int}$ are given
in Appendix \ref{sec:h_coll}.
${\cal H}_{rot}$ can be diagonalized in the space spanned
by the eigenstate $\ket{J, j_{3} ,J_{3};\, I, i_{3}, I_{3}}$,
where $J$ and $J_{3}$ ($I$ and $I_{3}$) are the eigenvalues
of the body fixed spin (isospin) operators and
$j_{3}$ ($i_{3}$) is the eigenvalue of the space fixed
spin (isospin) orator. The explicit representation of the eigenstate for the Euler angles
is given by the direct product of the Wigner \textit{D} functions
in real and isospin space \cite{rf:Braaten88}.

${\cal H}_{int}$ algebraically mixes together the space of the kaon and the collective rotation.  
By using the constraint Eq.~(\ref{eq:constraint}) with Eq.~(\ref{eq:isospin_rel}),
the basis vector for the whole space is given by
\begin{equation}
	\ket{J, j_{3}, J_{3}; I, i_{3}, -(J_{3}+I_{K3})}\,\ket{B(S, I_{K3})}.
\label{eq:whole_basis}
\end{equation}
Because the total Hamiltonian ${\cal H}$ commutes with $S$, the diagonalization is
well performed in the subspace with fixed eigenvalues $(J, j_{3})$, $(I, i_{3})$ and $S$.
However, since $J_{3}$,$I_{3}$, and $I_{K3}$ are not good quantum numbers individually,
the rotational band of combined system of the soliton and the kaon
is given by a linear combination of the basis Eq.~(\ref{eq:whole_basis}):
\begin{eqnarray}
	&&\ket{\Psi\left(S; J, j_{3}; I, i_{3}; \alpha\right)} =
	\sum_{J_{3},I_{K3}} C_{J_{3},I_{K3}}^{\Psi}\nonumber\\
	&&\times \ket{J, j_{3}, J_{3}; I, i_{3}, -(J_{3}+I_{K3})}\,\ket{B(S, I_{K3})},
\label{eq:whole_wf}
\end{eqnarray}
where $\alpha$ indicates quantum numbers other than $J$,$j_{3}$,$I$,$i_{3}$, and $S$.
This means that the isospin symmetry broken by the soliton (Sec.~\ref{sec:model})
would be restored by the collective rotation.
One of the scales for the isospin symmetry breaking is the expectation value of $\kappa_{3}$
for the rotational band:
\begin{equation}
	\kappa_{\Psi3} = \braketo{\Psi}{\kappa_{3}}{\Psi}.
\label{eq:kappa_p3}
\end{equation}
Of course, since we perform the quantization of the kaon and the collective rotation
around a specific Hartree state $\ket{B_{0}(S, I_{K3})}$ which breaks the isospin symmetry, 
the symmetry is not perfectly restored.
Thus, the calculation of the physical quantities with the rotational band Eq.~(\ref{eq:whole_wf})
is an estimate.

The number of bases in Eq.~(\ref{eq:whole_wf}) decreases from $(2J+1)(2I+1)$
due to the constraint Eq.~(\ref{eq:constraint}) and thus the axial symmetry of the soliton.
For example, the bases for the $\Sigma$ particle ($S=-1$,$J=1/2$,$I=1$) are given by
\begin{eqnarray}
	\left(\begin{array}{l}
		\ket{J, j_{3}, +1/2; I, i_{3}, -1}\,\ket{B(-1,+1/2} \\
		\ket{J, j_{3}, -1/2; I, i_{3},  0}\,\ket{B(-1,+1/2} \\
		\ket{J, j_{3}, +1/2; I, i_{3},  0}\,\ket{B(-1,-1/2} \\
		\ket{J, j_{3}, -1/2; I, i_{3}, +1}\,\ket{B(-1,-1/2}
	\end{array}\right).
\end{eqnarray}
After the diagonalization of the total Hamiltonian ${\cal H}$ with these bases,
we obtain the four rotational levels for $\Sigma$.
We assign the level with the lowest eigenvalue as $\Sigma$ in the real world.
As the shape of the chiral soliton approach the hedgehog one, 
other higher levels become heavy and disappear from the system.
For the other baryon, we equally do.

\section{Results\label{sec:results}}
For the effective action Eq.~(\ref{eq:action}),
we use Schwinger's proper time regularization \cite{rf:Ebert86,rf:Diakonov88,rf:Schwinger51}
with the cutoff parameter $\Lambda$.
And we assume that the valence quarks are in the lowest positive energy state.
Thus, we have four parameters: the dynamical quark mass $M$,
the current quark masses $m_{u} = m_{d}$ and $m_{s}$, and $\Lambda$.
We show in Table~\ref{tb:parameter} two parameter sets that we have considered.
The set (A) is the same as the parameters in Ref.~\cite{rf:akiyama03} and
used for comparison with the results.
The set (B) has been obtained by fitting to the experimental values.
\begin{table}
	\caption{The sets of parameters: 
	the dynamical quark mass $M$, the current quark masses $m_{u} = m_{d}$ and $m_{s}$,
	and the cutoff parameter $\Lambda$. All units are MeV.}
	\begin{ruledtabular}
	\begin{tabular}{ccccc}
	Set & $M$ & $m_{u}$ & $m_{s}$ & $\Lambda$\\
	\hline
	(A) & 400 & 6 & 200 & 700\\
	(B) & 400 & 6 & 190 & 700\\
	\end{tabular}
	\end{ruledtabular}
\label{tb:parameter}
\end{table}

\begin{figure}
	\includegraphics{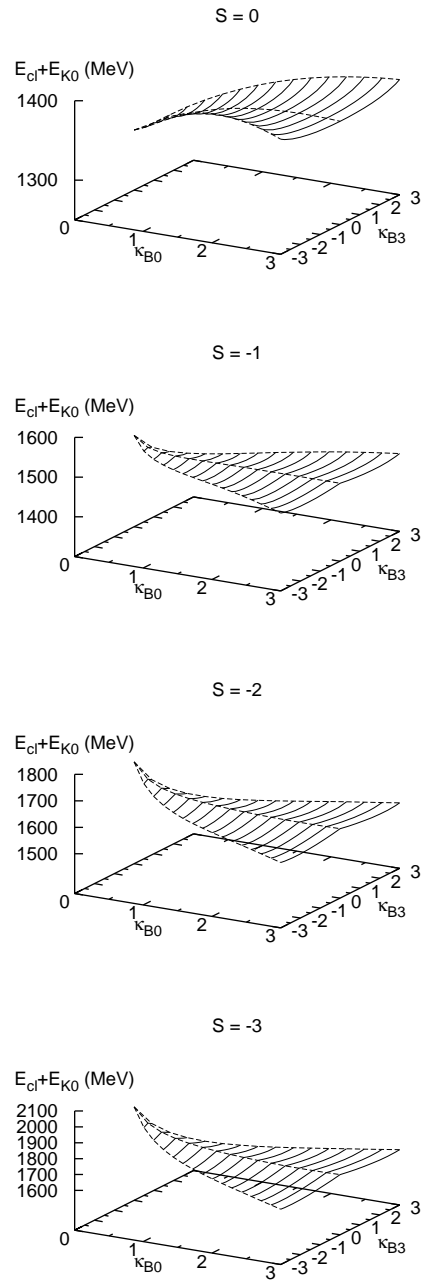}
	\caption{$E_{cl} + {E}_{K0}$, where ${E}_{K0}$ is the lowest eigenvalue
	of ${\cal H}_{K0}$ [Eq.~(\ref{eq:hK0_diag})].
	Here $\kappa_{B0}$ and $\kappa_{B3}$ are treated as parameters.}
\label{fig:e_k0}
\end{figure}

At first, we give the results for the set (A).
Let us denote the lowest eigenvalue of ${\cal H}_{K0}$ [Eq.~(\ref{eq:hK0_diag})] by $E_{K0}$. 
Then, $E_{cl}+{E}_{K0}$ is the lowest order approximation
of the energy for the soliton $+$ the kaon system.
This is a sufficient quantity to see the qualitative tendency of the system
in the $(\kappa_{B0},\kappa_{B3})$ space.
The more accurate calculation is performed below with the Hartree approximation.
Figure~\ref{fig:e_k0} shows the $\kappa_{B0}$ and $\kappa_{B3}$ dependence
of $E_{cl}+{E}_{K0}$ in the cases of $S = 0,-1,-2,-3$,
where $\kappa_{B0}$ and $\kappa_{B3}$ are treated as parameters.
As $\kappa_{B0}$ grows, the graphs decrease for $S = -1,-2,-3$.
It is because that the SU(3) symmetry breaking stabilizes the $\kappa_{B0} \not= 0$ state
for the strange sectors and the $\kappa_{B0} = 0$ state for the non-strange one.
On the other hand, the isospin symmetry is broken by the soliton solution.
Then the $\kappa_{B3} \not= 0$ states are stable for for $S = -1,-2,-3$
and the $\kappa_{B3} = 0$ state is stable for $S = 0$.
However, $E_{cl}+{E}_{K0}$ is an even function of $\kappa_{B3}$
because of the isospin symmetry of the model itself.

These graphs are flatter than our previous results \cite{rf:akiyama03}
which represent the behavior of the lowest eigenvalue of the $O(1)$ Hamiltonian
of $\cal H$ [Eq.~(\ref{eq:h})] in the large $N_{c}$ limit.
The difference is mainly due to the terms depending on $B_{3}$ of ${\cal H}_{K0}$.
These terms partially restore the isospin symmetry broken by the soliton in
the parameter space $(\kappa_{B0},\kappa_{B3})$. 
The flat regions in the graphs indicate that the Hamiltonian $E_{cl} + {\cal H}_{K0}$
accurately describes the combined system of kaon and soliton there.

Next, we show results of the Hartree approximation.
For $S = 0,-1,-2,-3$, the Hartree states which satisfy Eqs.~(\ref{eq:var_g}),
(\ref{eq:kappaB0}), and (\ref{eq:kappaB3}) are self-consistently searched.
Then, the values of $\kappa_{B0}$, $\kappa_{B3}$ are determined.
Table~\ref{tb:hartree} shows the resultant $\kappa_{B0}$, $|\kappa_{B3}|$,
$E_{cl}$, and $E_{B0}$ [Eq.~(\ref{eq:e_b})].
It is found from the process that the stable solutions of Eq.~(\ref{eq:var_g}) exist only
in the vicinity of $\kappa_{B3} \approx 0$ or $\kappa_{B0} \approx |\kappa_{B3}|$.
Moreover, Eqs.~(\ref{eq:kappaB0}) and (\ref{eq:kappaB3}) restrict the choice
to $\kappa_{B0} \approx |\kappa_{B3}|$.
In general, $\kappa_{B0} = |\kappa_{B3}|$ is not satisfied
by the dressed state Eq.~(\ref{eq:dressed}).
The only pure valence kaon state defined by Eq.~(\ref{eq:fock}) satisfies this relation.
Our result indicates that the kaon state is stable against the particle-antiparticle
creation in the vicinity of the solution.
As stated in Sec.~\ref{sec:coll_q}, the Hartree state does not correspond to
the absolute minimum of $E_{B0}(S, I_{K3})$,
but the energy difference is not more than $10$~MeV. 

The value of $\kappa_{B3}$ determines the shape of the chiral soliton \cite{rf:akiyama03}.
Since $\kappa_{B3} = 0$ for $S = 0$, the chiral soliton takes the hedgehog shape
in the Hartree approximation.
On the other hand, since $\kappa_{B3} \not= 0$ for $S \not= 0$,
the chiral soliton takes non-hedgehog shape.
Although, compared with our previous results \cite{rf:akiyama03}, the values of $\kappa_{B0}$
and $|\kappa_{B3}|$ are rather small, the conclusion about the shape of the chiral
soliton does not change.
From the present point of view, it is clear that the kaon states defined
in Ref.~\cite{rf:akiyama03} are unstable against the particle-antiparticle creation.
\begin{table}
	\caption{$\kappa_{B0}$, $|\kappa_{B3}|$, $E_{cl}$, and $E_{B0}$ [Eq.~(\ref{eq:e_b})]
	of the Hartree states for $S = 0, -1, -2, -3$. The set of parameters is (A).}
	\begin{ruledtabular}
	\begin{tabular}{ccccc}
	$S$ & $\kappa_{B0}$ & $|\kappa_{B3}|$ & $E_{cl}$~(MeV) & $E_{B0}$~(MeV)\\
	\hline
	0	& 0.00 & 0.00 & 1326 & 1326\\
	-1	& 0.81 & 0.81 & 1463 & 1487\\
	-2	& 1.64 & 1.64 & 1498 & 1600\\
	-3	& 2.52 & 2.52 & 1505 & 1713\\
	\end{tabular}
	\end{ruledtabular}
\label{tb:hartree}
\end{table}

In Table~\ref{tb:multiplet}, we show the expectation values of
$\kappa_{0}$, $\kappa_{3}$, and $E_{B}$ [Eq.~(\ref{eq:e_b})]
for each multiplet in the Hartree approximation.
Here, the only multiplets which contain the Hartree state $\ket{B_{0}(S,I_{K3})}$
with $\kappa_{B3} \leq 0$, are listed.
For $\kappa_{B3} \geq 0$, only the signs of $I_{K3}$ and $\braketo{B}{\kappa_{0}}{B}$ are reversed.
Because of the isospin symmetry breaking, the quantities in Table~\ref{tb:multiplet} take
different values even in a multiplet.
\begin{table}
	\caption{$\braketo{B}{\kappa_{0}}{B}$, $\braketo{B}{\kappa_{3}}{B}$,
	and $E_{B}$ [Eq.~(\ref{eq:e_b})] for the multiplet $\ket{B(S,I_{K3})}$
	in the Hartree approximation. $\ket{B_{0}}$ is the Hartree state corresponding to
	the lowest energy eigenvalue in a multiplet. The set of parameters is (A).}
	\begin{ruledtabular}
	\begin{tabular}{lccccc}
	$\ket{B(S,I_{K3})}$ & $\braketo{B}{\kappa_{0}}{B}$ & $\braketo{B}{\kappa_{3}}{B}$ &
	$E_{B}$~(MeV)\\
	\hline
	$\ket{B_{0}(0,0)}$	& 0.00	& 0.00	& 1326\\
	\hline
	$\ket{B_{0}(-1,+1/2)}$	& 0.81	& -0.81	& 1487\\
	$\ket{B(-1,-1/2)}$	& 0.68	& 0.68	& 1517\\
	\hline
	$\ket{B_{0}(-2,+1)}$	& 1.64 	& -1.64	& 1600\\
	$\ket{B(-2,0)}$		& 1.42 	& -0.22	& 1948\\
	$\ket{B(-2,-1)}$		& 1.20 	& 1.20	& 1671\\
	\hline
	$\ket{B_{0}(-3,+3/2)}$	& 2.52	& -2.52	& 1713\\
	$\ket{B(-3,+1/2)}$	& 2.27	& -1.10	& 2301\\
	$\ket{B(-3,-1/2)}$	& 2.02	& 0.33	& 2335\\
	$\ket{B(-3,-3/2)}$	& 1.76	& 1.76	& 1814\\
	\end{tabular}
	\end{ruledtabular}
\label{tb:multiplet}
\end{table}

Finally, we give the results by the collective rotation.
Table~\ref{tb:rot_band} shows the possible rotational bands by the diagonalization
with Eq.~(\ref{eq:whole_wf}).
As the deformations of the soliton become large,
the excitation energy between the lowest and the first excited state decrease. 
Table~\ref{tb:baryon} shows the lowest rotational band, which we assign as the baryon.
Here, $\kappa_{\Psi0} = \braketo{\Psi}{\kappa_{0}}{\Psi}$
similar to Eq.~(\ref{eq:kappa_p3}).

The calculated masses of $S = 0$ baryons ($N$, $\Delta$) are larger than the experimental value. 
It is pointed out in Ref.~\cite{rf:Weigel95,rf:Alkofer95} that these phenomena disappear 
by introducing the zero mode due to the hedgehog shape of the soliton.
For $S = 0$, we see $\braketo{B}{\kappa_{3}}{B} = 0$ in Table~\ref{tb:multiplet}.
It means that the soliton always takes the hedgehog shape.
Thus, the rotational band for $S = 0$ in Table~\ref{tb:baryon} also produce
$\kappa_{\Psi3} = 0$ exactly.

On the other hand, for $S \not= 0$ baryons ($\Lambda$, $\Sigma^{(*)}$, $\Xi^{(*)}$, $\Omega$),
the calculated masses are light.
We can recognize from a comparison between the Tables \ref{tb:hartree} and \ref{tb:baryon}, 
that it is due to the collective rotation in isospin and real space.
The rotational bands for $S \not= 0$ contain the fluctuation caused by the transition
within the multiplet $\ket{B(S, I_{K3})}$ in Table.~\ref{tb:multiplet}.
As a result we obtain the expectation value $\kappa_{\Psi3} \approx  0$ 
in Table~\ref{tb:baryon}.
\begin{table}
	\caption{The possible rotational bands with Eq.~(\ref{eq:whole_wf})
	for the set of parameters: (A).}
	\begin{ruledtabular}
	\begin{tabular}{cccccccc}
	S & Particle &		& Bands & & (MeV)\\
	\hline
	0	&	$N$		& 1382 \\
		& $\Delta$		& 1608 \\
	\hline
	-1	& $\Lambda$		& 1205 & 2448 \\
		& $\Sigma$		& 1239 & 2475 & 2691 & 5170 \\
		& $\Sigma^{*}$	& 1453 & 2478 & 2697 & 4951 & 5173 & 8888 \\
	\hline
	-2	& $\Xi$		& 1389 & 1848 & 2109 & 3023 \\
		& $\Xi^{*}$		& 1669 & 1848 & 2141 & 2764 & 3042 & 4400 \\
	\hline
	-3	& $\Omega$		& 1738 & 2124 & 2831 & 3901 \\
	\end{tabular}
	\end{ruledtabular}
\label{tb:rot_band}
\end{table}

\begin{table}
	\caption{$\kappa_{\Psi0}$, $|\kappa_{\Psi3}|$, Baryon masses $E_{\Psi}$.
	for the set of parameters: (A). Expt. denotes experimental value.}
	\begin{ruledtabular}
	\begin{tabular}{cccccc}
	S & Particle & $\kappa_{\Psi0}$ & $|\kappa_{\Psi3}|$ & $E_{\Psi}$~(MeV)  & Expt.~(MeV) \\
	\hline
	0	&	$N$		& 0.00 & 0.00 & 1382 & 939  \\
		& $\Delta$		& 0.00 & 0.00 & 1608 & 1232 \\
	\hline
	-1	& $\Lambda$		& 0.75 & 0.08 & 1205 & 1116 \\
		& $\Sigma$		& 0.75 & 0.08 & 1239 & 1193 \\
		& $\Sigma^{*}$	& 0.75 & 0.07 & 1453 & 1384 \\
	\hline
	-2	& $\Xi$		& 1.43 & 0.28 & 1389 & 1318 \\
		& $\Xi^{*}$		& 1.42 & 0.25 & 1669 & 1534 \\
	\hline
	-3	& $\Omega$		& 2.13 & 0.30 & 1738 & 1672 \\
	\end{tabular}
	\end{ruledtabular}
\label{tb:baryon}
\end{table}

Next, we show the possible rotational bands for the another set (B) of parameters
in Tables~\ref{tb:rot_bandB}.
\begin{table}
	\caption{The possible rotational bands with Eq.~(\ref{eq:whole_wf})
	for the set of parameters: (B).}
	\begin{ruledtabular}
	\begin{tabular}{cccccccc}
	S & Particle &		& Bands & & (MeV)\\
	\hline
	0	&	$N$		& 1382 \\
		& $\Delta$		& 1608 \\
	\hline
	-1	& $\Lambda$		& 1178 & 2499 \\
		& $\Sigma$		& 1212 & 2527 & 2742 & 5379 \\
		& $\Sigma^{*}$	& 1425 & 2529 & 2747 & 5159 & 5380 & 9332 \\
	\hline
	-2	& $\Xi$		& 1340 & 1869 & 2122 & 3177 \\
		& $\Xi^{*}$		& 1609 & 1868 & 2149 & 2925 & 3192 & 4763 \\
	\hline
	-3	& $\Omega$		& 1654 & 2093 & 2915 & 4157 \\
	\end{tabular}
	\end{ruledtabular}
\label{tb:rot_bandB}
\end{table}

\section{Summary and discussion\label{sec:summary}}
The stability of the hedgehog shape has been investigated
for the octet and decuplet baryons in the chiral quark soliton model.
We have expanded the collective Hamiltonian $\cal H$ up to the first order
in power of $1/N_{c}$ and the fluctuation ($\sigma_{\mu}$) around the mean field,
and separate it into the kaon,
the collective rotation, and the interaction Hamiltonian.

The kaon Hamiltonian are diagonalized by the Hartree approximation.
The resultant Hartree states describe the soliton and the kaon in the background soliton.
The kaons are almost in the valence states and stable against the particle-antiparticle creation. 
The shape of the soliton is controlled by the parameters $\kappa_{B0}$ and $\kappa_{B3}$
which also characterize the state of the kaon, and are self-consistently determined. 
The soliton takes the hedgehog shape for strangeness $S = 0$ and the non-hedgehog one
for $S = -1,-2,-3$.

The Hamiltonian for the collective rotation of the soliton is diagonalized
by the angular momentum basis in isospin and real space.
The interaction between the rotation and the soliton is treated by
the linear combination of the direct product of the angular momentum basis
and the Hartree state with the constraint due to the axial symmetry of the soliton.

The $S = 0$ rotational bands include only the pure hedgehog state and have the heavy masses.
Weigel \textit{et al.} investigated the quantum correction due to
the zero modes of the "hedgehog" soliton \cite{rf:Weigel95,rf:Alkofer95}.
The correction gives a large negative contribution ($\sim -400$~MeV) to the $N$ and $\Delta$ masses,
and their results are in good agreement with the experimental values.
Their analysis is valid also for the $S = 0$ baryons in our approach.

On the other hand, for $S \not= 0$, the rotation mixes the Hartree states
with nonzero values of $\kappa_{B3}$, and the resultant rotational bands
consist of the mixing of the "non-hedgehog" states.
As a result, the $S \not= 0$ baryons have light masses.
It is clear from the separation of the variables in the collective Hamiltonian
that the light masses of the $S \not= 0$ rotational bands are due to the rotation
of the non-hedgehog soliton.
The calculation of Weigel \textit{et al.} would be inapplicable to the $S \not= 0$ cases,
because the soliton take the the "non-hedgehog" shape and the zero mode does not arise.
Thus, the consistent treatment of the deformation of the soliton and the zero mode 
would resolve the reversed mass order between the $S \not= 0$ and $S = 0$ baryons. 

The multiplicity of the basis due to the axial symmetric deformation of the soliton necessarily
produces the excited states with the same spin-flavor quantum numbers as the ground state.
For $S = -1$ baryon, the energy differences between the ground state and the first excited one
are too large ($\sim 1$~Gev) because of the small deformation of the soliton.
Thus, these excited states may be beyond the validity of the model.
On the other hand, the first excited states for $S = -2, -3$ have the moderate mass
and would be assigned to the excited baryon in the real world.
However, the strict assignment needs more symmetric treatment
of the Hartree states and the collective quantization. 

Our study originated from the observation \cite{rf:akiyama03} that
the $S = -1, -2$ baryons consist of the quarks with different masses and
the inertial force would deform the shape of the soliton. 
In fact, if SU(3) symmetry breaking vanishes ($m_{8} = 0$) in the quark Hamiltonian,
the soliton always takes the hedgehog shape independently of the strangeness. 
Thus, it is surprising that $\Omega$ ($\ket{sss}$) also takes the non-hedgehog shape.
The solution to the problem is in Eqs.~(\ref{eq:isospin_rel}) and (\ref{eq:rot_hq0}).
The former indicates that the emission and absorption of the kaon
cause the isospin rotation of the soliton.
The latter indicates that its recoil effects (the rotated quark mass) asymmetrically act
on the isodoublet quarks in the body fixed frame and cause the direction of the isospin
vector the asymmetry.
Since the soliton consists of the valence and sea quarks,
even though the masses of valence quarks are equivalent, 
these processes happen as far as the kaon states bear the isospin.
As a result, the soliton takes the non-hedgehog shape for $S = -1,-2,-3$, stably.

Inclusion of the locality is in progress but a complicated task.
Therefore we will include the results in the future work.

\begin{acknowledgments}
One of us (SA) would like to thank H.~Kondo and C.~Yasuda
for many helpful discussions and encouragement.
\end{acknowledgments}

\appendix
\section{Functions related to the rotation into the strange direction\label{sec:rot_s}}
At first, we define the following quantities:
\begin{eqnarray*}
	f &=& \frac{1-\cos \sqrt{\kappa_{0}}}{\kappa_{0}},\\
	g &=& \frac{\sin \sqrt{\kappa_{0}}}{\sqrt{\kappa_{0}}},\\
	\chi_{+} &=& \left(\begin{array}{c}1\\0\end{array}\right),
	\chi_{-} = \left(\begin{array}{c}0\\1\end{array}\right),\\
	\tilde{\tau}_{j} &=& \tau_{j}-\kappa_{j} f\hspace{1cm}(j = 1,2,3).
\end{eqnarray*}
The Wigner \textit{D} function $D^{(8)}_{\mu\nu}(A_{s})$  [Eq.~(\ref{eq:d_func})] 
are explicitly given as
\begin{eqnarray}
	D^{(8)}_{ij} &=& \delta_{ij} (1-\kappa_{0} f) 
			+ \frac{1}{2} \kappa_{i} \kappa_{j} f^{2},\\
	D^{(8)}_{4j} &=& \frac{-i g}{\sqrt{2}} \left(
			D^{\dagger} \tilde{\tau}_{j} \chi_{+}
			-\chi_{+}^{\dagger} \tilde{\tau}_{j} D
	\right),\\
	D^{(8)}_{5j} &=& \frac{- g}{\sqrt{2}} \left(
			D^{\dagger} \tilde{\tau}_{j} \chi_{+}
			+\chi_{+}^{\dagger} \tilde{\tau}_{j} D
	\right),\\
	D^{(8)}_{6j} &=& \frac{-i g}{\sqrt{2}} \left(
			D^{\dagger} \tilde{\tau}_{j} \chi_{-}
			-\chi_{-}^{\dagger} \tilde{\tau}_{j} D
	\right),\\
	D^{(8)}_{7j} &=& \frac{- g}{\sqrt{2}} \left(
			D^{\dagger} \tilde{\tau}_{j} \chi_{-}
			+\chi_{-}^{\dagger} \tilde{\tau}_{j} D
	\right),\\
	D^{(8)}_{8j} &=& -\frac{\sqrt{3}}{2} g^{2} \kappa_{j},\\
	D^{(8)}_{84} &=& -\sqrt{\frac{3}{2}} g (1-\kappa_{0} f) i(D^{\dagger}_{1}-D_{1}),\\	
	D^{(8)}_{85} &=& -\sqrt{\frac{3}{2}} g (1-\kappa_{0} f) (D^{\dagger}_{1}+D_{1}),\\	
	D^{(8)}_{86} &=& -\sqrt{\frac{3}{2}} g (1-\kappa_{0} f) i(D^{\dagger}_{2}-D_{2}),\\	
	D^{(8)}_{87} &=& -\sqrt{\frac{3}{2}} g (1-\kappa_{0} f) (D^{\dagger}_{1}+D_{1}),\\
	D^{(8)}_{88} &=& 1-\frac{3}{2} g^{2} \kappa_{0},
\end{eqnarray}
where $i,j = 1,2,3$.

The isodoublet spinor $C_{\mu}$ [Eq.~(\ref{eq:c_cof})] are represented as
\begin{widetext}
\begin{eqnarray}
	C_{j} &=& -i f \left(2 \tau_{j} - \kappa_{j} f\right) D,\\
	C_{4} &=& \sqrt{2} g \left[
		\chi_{+} + \frac{1-g}{\kappa_{0} g} (D^{\dagger}_{1}+D_{1}) D
		-f (D^{\dagger}_{1}-D_{1}) D
	\right],\\
	C_{5} &=& -i\sqrt{2} g \left[
		\chi_{+} + \frac{1-g}{\kappa_{0} g} (D^{\dagger}_{1}-D_{1}) D
		-f (D^{\dagger}_{1}+D_{1}) D
	\right],\\
	C_{6} &=& \sqrt{2} g \left[
		\chi_{-} + \frac{1-g}{\kappa_{0} g} (D^{\dagger}_{2}+D_{2}) D
		-f (D^{\dagger}_{2}-D_{2}) D
	\right],\\
	C_{7} &=& -i\sqrt{2} g \left[
		\chi_{-} + \frac{1-g}{\kappa_{0} g} (D^{\dagger}_{2}-D_{2}) D
		-f (D^{\dagger}_{2}+D_{2}) D
	\right],\\
	C_{8} &=& -i \sqrt{3} g^{2} D.
\end{eqnarray}

The functions $N_{i}$, $F_{i}$, $G_{i}$ ($i = 1,2$) in Eq.~(\ref{eq:kaon_rel}) are defined by
\begin{eqnarray}
	N_{1} &=& \left(D^{(8)}_{84}\right)^{2}+\left(D^{(8)}_{85}\right)^{2},\hspace{5mm}
	N_{2} = \left(D^{(8)}_{86}\right)^{2}+\left(D^{(8)}_{87}\right)^{2},\nonumber\\
	F_{1} &=& \frac{i}{2} \left[
		(1+\kappa_{3} f) \left(D^{\dagger}\tilde{P}-\tilde{P}^{\dagger}D \right)
		+(1-\kappa_{0} f) \left(D^{\dagger}\tau_{3}\tilde{P}-\tilde{P}^{\dagger}\tau_{3}D \right)
		\right],\nonumber\\
	F_{2} &=& \frac{i}{2} \left[
		(1-\kappa_{3} f) \left(D^{\dagger}\tilde{P}-\tilde{P}^{\dagger}D \right)
		-(1-\kappa_{0} f) \left(D^{\dagger}\tau_{3}\tilde{P}-\tilde{P}^{\dagger}\tau_{3}D \right)
		\right],\nonumber\\	
	G_{1} &=& \frac{1}{2} (1-\kappa_{0} f) \left[
			\left(g-\frac{(1-g)}{\kappa_{0}}\kappa_{3}\right)
				\left(D^{\dagger}\tilde{P}+\tilde{P}^{\dagger}D \right)
			+\left(D^{\dagger}\tau_{3}\tilde{P}+\tilde{P}^{\dagger}\tau_{3}D \right)
		\right],\nonumber\\
	G_{2} &=& \frac{1}{2} (1-\kappa_{0} f) \left[
			\left(g+\frac{(1-g)}{\kappa_{0}}\kappa_{3}\right)
				\left(D^{\dagger}\tilde{P}+\tilde{P}^{\dagger}D \right)
			-\left(D^{\dagger}\tau_{3}\tilde{P}+\tilde{P}^{\dagger}\tau_{3}D \right)
		\right].
\end{eqnarray}

\section{Collective Hamiltonian\label{sec:h_coll}}
We define following quantities:
\begin{eqnarray}
	A_{Kj} &=& \frac{1}{2} \left(D^{\dagger} \tau_{j} P + P^{\dagger} \tau_{j} D\right)
	\hspace{1cm}(j = 1,2,3),\nonumber\\
	g_{B} &=& \frac{\sin \sqrt{\kappa_{B0}}}{\sqrt{\kappa_{B0}}},\hspace{5mm}
	\frac{1}{\Phi_{(\pm)}} = \frac{1}{2}
		\left(\frac{1}{U_{44}} \pm \frac{1}{U_{66}}\right),\hspace{5mm}
	\Delta_{(\pm)} = \frac{1}{2}
		\left(\frac{\Delta_{44}}{U_{44}} \pm \frac{\Delta_{66}}{U_{66}}\right), \nonumber\\
	E_{ind} &=& 3 m_{8} g_{B}^{2} \left[
			\kappa_{B0} \left(\Gamma_{8}-\frac{B_{3}\Delta_{38}}{U_{33}}\right)
		+ \frac{\kappa_{B3}}{\sqrt{3}}
				\left(\Gamma_{3}-\frac{B_{3}\Delta_{33}}{U_{33}}\right)
			\right]
		+\frac{B_{3}^{2}}{2 U_{33}}. \nonumber
\end{eqnarray}
Furthermore, for the arbitrary three dimensional vectors ${\bf x} = (x_{1},x_{2},x_{3})$
and ${\bf y} = (y_{1},y_{2},y_{3})$,
we introduce following symbol :
\begin{equation*}
	\left({\bf x} \cdot {\bf y}\right)_{\perp} = {\bf x} \cdot {\bf y} - x_{3} y_{3}.
\end{equation*}

The collective Hamiltonian ${\cal H}$ [Eq.~(\ref{eq:h})] is expanded up to the first order
in power of $1/N_{c}$ and $\sigma_{\mu}$  and separated as
\begin{equation}
	{\cal H} = E_{cl} + {\cal H}_{K} + {\cal H}_{rot} + {\cal H}_{int},
\end{equation}
where $E_{cl}$ is the classical soliton energy,
${\cal H}_{K}$ is the part containing only the kaon operators $D$ and $P$ and
describes the kaon in the background soliton,
${\cal H}_{rot}$ is the part containing only the angular momenta $I_{j}$ and $J_{j}$ and
represents the collective rotation of the system
in isospin and real space, and ${\cal H}_{int}$ represents
an interaction between the kaon and the rotation. 

The kaon Hamiltonian ${\cal H}_{K}$ is given by
\begin{eqnarray}
	{\cal H}_{K} &=& {\cal H}_{K0} + {\cal H}_{K1},
\label{eq:hK}\\
	{\cal H}_{K0} &=& E_{ind}
		+\frac{1}{4 \Phi_{(+)}} P^{\dagger} P
		+\frac{1}{4 \Phi_{(-)}} P^{\dagger} \tau_{3} P\nonumber\\
		&&+ 3 \left[
			\frac{B_{8}^{2} }{8 \Phi_{(+)}} 
			+ m_{8} \left(\Delta_{(+)} B_{8}-\Gamma_{8}
			+ \frac{B_{3}\Delta_{38}}{U_{33}}\right)
		\right] \kappa_{0}
		+ \sqrt{3} \left(\frac{B_{8}}{4 \Phi_{(+)}} + m_{8} \Delta_{(+)}\right) S \nonumber\\
		&&+ 3 \left[
			\frac{B_{8}^{2}}{8 \Phi_{(-)}} 
			+ m_{8} \left(\Delta_{(-)} B_{8}-\frac{\Gamma_{3}}{\sqrt{3}}
			+\frac{B_{3}\Delta_{33}}{\sqrt{3}U_{33}}\right)
		\right] \kappa_{3}
		+ \sqrt{3} \left(\frac{B_{8}}{4 \Phi_{(-)}} + m_{8} \Delta_{(-)}\right) 2 I_{K3},
\label{eq:hK0}\\
	{\cal H}_{K1} &=& \frac{1}{2(U_{11}V_{11}-W_{11}^{2})}
		\left[
			V_{11} \innerP{{\bf I}_{K}}{{\bf I}_{K}}
			+ 2 \sqrt{3} m_{8} \left(V_{11}\Delta_{11}-W_{11}{\tilde \Delta}_{11} \right)
				\innerP{{\bm \kappa}}{{\bf I}_{K}}
		\right]\nonumber\\
 		&&+ \frac{1}{\Phi_{(+)}}\left(\frac{S^2}{8}-\frac{1}{3}{\bf I}_{K}^{2}\right)
		+\frac{I_{K3}^{2}}{2 \Phi_{(+)}} 
		+ \frac{1}{3 \Phi_{(-)}} \left(
			S I_{K3} - \left[{\bf I}_{K} \times {\bf A}_{K}\right]_{3}
		\right)\nonumber\\
		&&+ \frac{\sqrt{3}}{2} \left(\frac{B_{8}}{4\Phi_{(-)}}+m_{8} \Delta_{(-)}\right) \kappa_{3} S
		+ \sqrt{3} \left(\frac{B_{8}}{4 \Phi_{(+)}} + m_{8} \Delta_{(+)}\right)
			\kappa_{3} I_{K3}\nonumber\\
		&&+ \frac{\sqrt{3}}{2} \left(\frac{B_{8}}{4\Phi_{(+)}}-m_{8} \Delta_{(+)}\right) \kappa_{0} S
		+ \left[
			\frac{B_{8}^{2}}{4 \Phi_{(+)}}
			- m_{8} \left(
				\Delta_{(+)} B_{8}-\Gamma_{8} + \frac{B_{3}\Delta_{38}}{U_{33}}
			\right)
		\right] \kappa_{0}^{2}\nonumber\\
		&&+ \sqrt{3} \left(\frac{B_{8}}{4\Phi_{(-)}}-m_{8} \Delta_{(-)}\right) \kappa_{0} I_{K3}
		+ \left[
			\frac{B_{8}^{2}}{4 \Phi_{(-)}}
			- m_{8} \left(\Delta_{(-)} B_{8}-\frac{\Gamma_{3}}{\sqrt{3}}
			+\frac{B_{3}\Delta_{33}}{\sqrt{3} U_{33}}\right)
		\right] \kappa_{0} \kappa_{3},
\label{eq:hK1}
\end{eqnarray}
where ${\cal H}_{K0}$ is the bilinear part of ${\cal H}_{K}$ in power of $D$ and $P$
and ${\cal H}_{K1}$ is the higher order one.

The collective rotation Hamiltonian ${\cal H}_{rot}$
and the kaon-rotation interaction Hamiltonian ${\cal H}_{int}$ are given by
\begin{eqnarray}
	{\cal H}_{rot} &=& \frac{1}{2(U_{11}V_{11}-W_{11}^{2})}
		\left[
			V_{11} \innerP{{\bf I}}{{\bf I}}
			+ U_{11} \innerP{{\bf J}}{{\bf J}}
			+ 2 W_{11} \innerP{{\bf I}}{{\bf J}}
		\right]\nonumber\\
		&&+ \frac{1}{2 U_{33}} J_{3}^{2}
		+ \frac{J_{3}}{U_{33}} \left[
			3m_{8} g_{R}^2 \left(
				\kappa_{B0} \Delta_{38} + \frac{\kappa_{B3}}{\sqrt{3}} \Delta_{33}
			\right)-B_{3}
		\right],
\label{eq:hrot}\\
		{\cal H}_{int} &=& -\frac{1}{2 \Phi_{(+)}} \innerP{{\bf I}_{K}}{{\bf I}} 
		- \sqrt{3} \left(\frac{B_{8}}{4 \Phi_{(+)}} + m_{8} \Delta_{(+)}\right)
			\innerP{{\bm \kappa}}{{\bf I}}
		-\frac{1}{2 \Phi_{(-)}} \left[{\bf I} \times {\bf A}_{K}\right]_{3}
		\nonumber\\
		&& + \frac{1}{(U_{11}V_{11}-W_{11}^{2})}
		\left[
			V_{11} \innerP{{\bf I}_{K}}{{\bf I}}
			+ \sqrt{3} m_{8} \left(
				V_{11}\Delta_{11}-W_{11}{\tilde \Delta}_{11}
			 \right) \innerP{{\bm \kappa}}{{\bf I}} \right. \nonumber\\
			&&+  \left. W_{11} \innerP{{\bf I}_{K}}{{\bf J}}
			+ \sqrt{3} m_{8} \left(
				W_{11}\Delta_{11}-U_{11}{\tilde \Delta}_{11}
			\right) \innerP{{\bm \kappa}}{{\bf J}}
		\right]\nonumber\\
		&&+ J_{3} \left[
			\frac{S}{4 \Phi_{(-)}}
			+\frac{I_{K3}}{2 \Phi_{(+)}} 
			+ \sqrt{3} \left(
				\frac{B_{8}}{4 \Phi_{(+)}} + m_{8} \Delta_{(+)}
			\right) \kappa_{3}
			+ \sqrt{3} \left(
				\frac{B_{8}}{4 \Phi_{(-)}} + m_{8} \Delta_{(-)}
			\right) \kappa_{0} \right. \nonumber\\
			&&- \left. \frac{3 m_{8}}{U_{33}} \left(
				\kappa_{0} \Delta_{38} + \frac{\kappa_{3}}{\sqrt{3}} \Delta_{33}
			\right)
		\right].
\label{eq:hint}
\end{eqnarray}
\end{widetext}


\end{document}